\documentclass[12pt]{article}
\usepackage[T1]{fontenc}
\usepackage{graphicx}

\usepackage{enumitem}

\newcommand{\Figure}{Figure\xspace}
\newcommand{\Algorithm}{Algorithm\xspace}
\newcommand{\Algorithms}{Algorithms\xspace}
\newcommand{\SectionAbv}{Section\xspace}

\usepackage{amsmath,amssymb,latexsym,amsopn,amsfonts}
\usepackage[boxed,ruled,vlined,linesnumbered,noresetcount]{algorithm2e}

\usepackage{fullpage}
\usepackage{framed} 
\usepackage[normalem]{ulem}
\usepackage{array,xspace}
\usepackage{graphicx}
\usepackage{color}
\usepackage[dvipsnames]{xcolor}
\definecolor{orange}{rgb}{0.99, 0.27, 0.0}
\definecolor{frenchBlue}{rgb}{0.13, 0.55, 0.13}
\newcommand{\ems}[1]{\textcolor{black}{#1}}
\newcommand{\emsA}[1]{\textcolor{black}{#1}}
\newcommand{\algSize}{normalsize} 

\newcommand{\blitza}{{\usefont{U}{ulsy}{m}{n}\symbol{'011}}}

\newcommand{\B}{}
\newcommand{\BB}{}
\newcommand{\BBB}{}

\newcommand{\FF}{\vspace*{\medskipamount}}

\usepackage{wrapfig}

\newcommand{\bigO}{\mathcal{O}\xspace}
\newcommand{\remove}[1]{}
\newcommand{\reduce}[1]{#1}

\newcommand{\done}{\mathsf{result}\xspace}

\newcommand{\Correct}{\mathit{Correct}\xspace}

\newcommand{\etal}{\emph{et al.}\xspace}
\newcommand{\eg}{\emph{e.g.,}\xspace}

\newcommand{\ie}{\emph{i.e.,}\xspace}
\newcommand{\Ie}{\emph{I.e.,}\xspace}

\newtheorem{theorem}{Theorem}[section]
\newtheorem{lemma}[theorem]{Lemma}
\newtheorem{definition}{Definition}[section]

\newtheorem{assumption}{Assumption}
\newtheorem{corollary}[theorem]{Corollary}
\newtheorem{myClaim}{Claim}
\newcommand{\true}{\mathsf{True}\xspace}

\newcommand{\false}{\mathsf{False}\xspace}

\newcommand{\sP}{\mathcal{P}\xspace}

\SetKwInOut{Input}{input}
\SetKwInOut{Output}{output}

\newenvironment{claimProof}{\par\noindent\textbf{Proof of Claim  \clmcnt\space}}{\hfill $\Box_{Claim ~ \clmcnt}$}
\newenvironment{claimProofSketch}{\par\noindent\textbf{Proof Sketch of Claim  \clmcnt\space}}{\hfill $\Box_{Claim ~ \clmcnt}$}
\newenvironment{lemmaProof}{\par\noindent\textbf{Proof of Lemma  \lemcnt\space}}{\hfill $\Box_{Lemma ~ \lemcnt}$}
\newenvironment{lemmaProofSketch}{\par\noindent\textbf{Proof Sketch of Lemma  \lemcnt\space}}{\hfill $\Box_{Lemma ~ \lemcnt}$}
\newenvironment{theoremProof}{\par\noindent\textbf{Proof of Theorem  \thmcnt\space}}{\hfill $\Box_{Theorem ~ \thmcnt}$}
\newenvironment{theoremProofSketch}{\par\noindent\textbf{Proof Sketch of Theorem  \thmcnt\space}}{\hfill $\Box_{Theorem ~ \thmcnt}$}
\newcommand{\clmcnt}{0}
\newcommand{\lemcnt}{0}
\newcommand{\thmcnt}{0}

\newcommand{\Section}[1]{\BBB\section{#1}\BB}

\newcommand{\Subsection}[1]{\smallskip\noindent \textbf{#1.}~~~}
\newcommand{\Subsubsection}[1]{\noindent \emph{#1.}~~}

\usepackage[utf8]{inputenc}

\newcommand{\technicalReport}[1]{#1}
\newcommand{\extendedAbstract}[1]{}

\begin{document}

\technicalReport{\title{Self-stabilizing Byzantine-tolerant Recycling \footnote{This document is a complementary technical report to the extended abstract version of this work~\cite{TR}.}}}

\extendedAbstract{\title{Self-stabilizing Byzantine-tolerant Recycling}}

	\author{Chryssis Georgiou \footnote{ Computer Science, Univ. Cyprus, Cyprus.
				Email: \texttt{chryssis@cs.ucy.ac.cy}} \and
	Michel Raynal \footnote{ IRISA, Univ. Rennes 1, France.
			Email: \texttt{michel.raynal@irisa.fr}} \and
	Elad M.\ Schiller \footnote{ Computer Science and Engineering, Chalmers Univ. Tech., Sweden.
			Email: \texttt{elad@chalmers.se}}}
%


\maketitle              

\begin{abstract}
Numerous distributed applications, such as cloud computing and distributed ledgers, necessitate the system to invoke asynchronous consensus objects an unbounded number of times, where the completion of one consensus instance is followed by the invocation of another. With only a constant number of objects available, object reuse becomes vital. 

We investigate the challenge of object recycling in the presence of \emph{Byzantine} processes, which can deviate from the algorithm code in any manner. Our solution must also be \emph{self-stabilizing}, as it is a powerful notion of fault tolerance. Self-stabilizing systems can recover automatically after the occurrence of \emph{arbitrary transient-faults}, in addition to tolerating communication and (Byzantine or crash) process failures, provided the algorithm code remains intact.

We provide a recycling mechanism for asynchronous objects that enables their reuse once their task has ended, and all non-faulty processes have retrieved the decided values. This mechanism relies on synchrony assumptions and builds on a new self-stabilizing Byzantine-tolerant synchronous multivalued consensus algorithm, along with a novel composition of existing techniques.
\end{abstract}


\Section{Introduction}
\label{sec:intro}
\ems{We study the problem of recycling asynchronous consensus objects.
We propose a more robust solution than the state-of-the-art solution to achieve this goal.
}

\Subsection{Fault model} 
\label{sec:fdIn}
We study solutions for message-passing systems. 
We model a broad set of failures that can occur to computers and networks. 
\extendedAbstract{Our}\technicalReport{Specifically, our} model includes up to $t$ process failures, \ie crashed or Byzantine~\cite{DBLP:journals/toplas/LamportSP82}. \ems{In detail, the adversary completely controls any Byzantine node, \eg the adversary can send a fake message that the node never sent, modify the payload of its messages, delay the delivery of its messages, or omit any subset of them.} 
\technicalReport{The adversary can challenge the algorithm by creating failure patterns in which a fault occurrence appears differently to different system components.
The adversary is empowered with the unlimited ability to coordinate the most severe failure patterns.} 
We assume a known maximum number, $t$, of \ems{Byzantine} processes. 
\ems{For solvability's sake, we} also restrict the adversary from letting a \ems{Byzantine} process impersonate a non-faulty one, \ems{\ie as in~\cite{DBLP:books/sp/Raynal18}, we assume private channels between any pair of nodes.} 


\Subsection{Self-stabilization} 
\label{sec:SelfStabIntro}
In addition to the failures captured by our model, we also aim to recover from \emph{arbitrary transient-faults}, \ie any temporary violation of assumptions according to which the system was designed to operate. This includes the corruption of control variables, such as the program counter, packet payload, and indices, \eg sequence numbers, which are responsible for the correct operation of the studied system, as well as operational assumptions, such as that at least a distinguished majority of processes never fail. Since the occurrence of these failures can be arbitrarily combined, we assume that these transient-faults can alter the system state in unpredictable ways. In particular, when modeling the system, Dijkstra~\cite{DBLP:journals/cacm/Dijkstra74} assumes that these violations bring the system to an arbitrary state from which a \emph{self-stabilizing system} should recover. Dijkstra requires recovery after the last occurrence of a transient-fault and once the system has recovered, it must never violate the task \ems{specifications. \Ie there could be any finite number of transient faults before the last one occurs, which may leave the system in an arbitrary state. Moreover, recovery from an arbitrary system state is demonstrated once all transient faults cease to happen, see~\cite{DBLP:series/synthesis/2019Altisen,DBLP:books/mit/Dolev2000}\technicalReport{ for details}.} 

\Subsubsection{Memory constraints} 
\ems{In the absence of transient faults, one can safely assume that the algorithm variables, such as a counter for the message sequence number, are unbounded. 
This assumption can be made valid for any practical setting by letting each counter use enough bits, say, $64$, because counting (using sequential steps) from zero to the maximum value of the counter will take longer than the time during which the system is required to remain operational. 
Specifically, if each message transmission requires at least one nanosecond, it would take at least 584 years until the maximum value can be reached. 
However, in the context of self-stabilization, a single transient fault can set the counter value into one that is close to the maximum value.
Thus, any self-stabilizing solution must cope with this challenge and use only bounded memory and communication.}

\Subsubsection{\ems{Self-stabilization via algorithmic transformation}} 
\ems{This work is dedicated to designing a generic transformer, which takes an algorithm as input and systematically redesigns it into its self-stabilizing variation as output. Existing transformers differ in the range of input algorithms and fault models they can transform, see Dolev~\cite[2.8]{DBLP:books/mit/Dolev2000}, Katz and Perry~\cite{DBLP:conf/podc/KatzP90}, Afek \etal~\cite{DBLP:conf/wdag/AfekKY90}, and Awerbuch \etal~\cite{DBLP:conf/wdag/AwerbuchPVD94}.}

\ems{Dolev, Petig, and Schiller~\cite{DBLP:journals/algorithmica/DolevPS23} (DPS in short) proposed a transformer of crash-tolerant algorithms for asynchronous message-passing systems into ones that also recover from transient faults\technicalReport{ via self-stabilization}.}
\ems{\technicalReport{Georgiou \etal~\cite{DBLP:conf/netys/GeorgiouGLS19} implemented DPS.}} 
\ems{Lundström \etal show DPS's applicability to various communication abstractions, such as atomic snapshot~\cite{DBLP:conf/podc/GeorgiouLS19}, consensus~\cite{DBLP:conf/icdcn/LundstromRS21,DBLP:conf/edcc/LundstromRS21}, reliable broadcast~\cite{DBLP:conf/netys/LundstromRS20}, and state-machine replication~\cite{DBLP:conf/sss/LundstromRS22}.}
\ems{DPS mandates that (DPS.i) the input algorithm guarantees,  after the last transient fault occurrence, the completion of each invocation of the communication abstraction, \ie it should eventually terminate regardless of the starting state. This condition facilitates the eventual release of resources used by each invocation.}
\ems{Additionally, (DPS.ii) it associates a sequence number with each invocation to differentiate the resources utilized by different invocations. This enables the recycling of resources associated with obsolete invocations through a sliding window technique, along with a global restart once the maximum sequence number is reached.}



\ems{Recently, DPS was utilized by Duvignau \etal~\cite{DBLP:conf/sss/DuvignauRS22} for converting Byzantine fault-tolerant (BFT) reliable broadcast proposed by Bracha and Toueg~\cite{DBLP:conf/podc/BrachaT83} into a Self-Stabilizing BFT (SSBFT) variation.
This solution recycles reliable broadcast objects using synchrony assumptions.
It also relies on the fact that the process may allocate independent local memory and sequence numbers per sender.
However, consensus objects often use shared sequence numbers, and thus, parts of their local memories are codependent.
Therefore, we use another approach.}

%

\Subsection{\ems{Problem description}} 
\ems{This work studies an important building block that is needed for the SSBFT implementation of asynchronous consensus objects.
With only a bounded number of consensus objects available, it becomes essential to reuse them robustly. 
We examine the case in which the repeated invocation of consensus needs to reuse the same memory space and the $(k+1)$-th invocation can only start after the completion of the $k$-th instance. 
In an asynchronous system that uses only a bounded number of objects, ensuring the termination of the $k$-th instance before invoking the $(k+1)$-th might be crucial, \eg for total order broadcasting, as in some blockchains.
Thus, we require SSBFT consensus objects to eventually terminate regardless of their starting state, as in (DPS.i).}


We focus \ems{on addressing the challenge of recycling asynchronous consensus objects after they have completed their task and delivered their decision to all non-faulty processes. 
This task becomes complex due to the presence of asynchrony and Byzantine failures. 
Utilizing the joint sequence numbers of (DPS.ii) for recycling consensus objects is not straightforward, because it requires ensuring that all non-faulty processes have delivered the decided value (for the $k$-th consensus invocation) as well as agreeing that such collective delivery occurred before incrementing the sequence number counter (that is going to be used by the $(k+1)$-th invocation). 
To overcome this chicken-and-egg problem, we relax the problem requirements by allowing the recycling mechanism to depend on synchrony assumptions.}
To mitigate the impact of these assumptions, a single recycling action can be performed for a batch of $\delta$ objects, where $\delta$ is a predefined constant determined by the available memory. 
Thus, our approach \ems{facilitates asynchronous networking} in communication-intensive components, \ems{\ie the consensus objects,} while \emsA{synchronous recycling} \ems{actions are performed according to a load parameter,} $\delta$.

\Subsection{\ems{Our solution in a nutshell}}
\ems{Our solution aims to emulate (DPS.ii) by incorporating synchrony assumptions specifically for the recycling service, while keeping the consensus object asynchronous to handle intensive message exchange.}

\ems{To begin, we maintain an index that points to the most recently invoked object in a constant-size array.
In order to ensure that all non-faulty processes agree on the value of the index, we utilize a novel technique called \emph{simultaneous increment-or-get indexes} (SGI-index).
When recycling an object, we increment the index, but this increment is performed only after an agreement among the non-faulty processes that the relevant object has made its decision and delivered it to all non-faulty processes.
Thus, we use a new SSBFT multivalued consensus before each increment, ensuring that consensus is reached before the increment. \Ie all needed deliveries had occurred before the increment.}

\ems{Additionally, our solution answers how an SSBFT asynchronous consensus object can provide an indication that at least one non-faulty process has made a decision and delivered. We utilize this indication as input to trigger the recycling action, effectively incorporating it into the SSBFT multivalued consensus.}

\Subsection{Related work}
%
%
\reduce{\Subsubsection{Object recycling}} 
Object recycling was studied mainly in the context of \ems{crash-tolerant (non-BFT)} systems~\cite{DBLP:conf/iwmm/PlainfosseS95,DBLP:conf/ipps/VeigaF05}.
There are a few \ems{(non-self-stabilizing)} implementations of garbage collection in the presence of Byzantine processes, \eg~\cite{DBLP:conf/opodis/OliveiraMB16}.

\reduce{
\Subsubsection{Impossibilities} 
\label{sec:impo}
As mentioned, FLP~\cite{DBLP:journals/jacm/FischerLP85} concluded that consensus is impossible to solve deterministically in asynchronous settings in the presence of even a single crash failure. In~\cite{DBLP:journals/ipl/FischerL82} it was shown that a lower bound of $t+1$ communication steps are required to solve consensus deterministically in both synchronous and asynchronous environments. 
%
%
In the presence of Byzantine faults, the consensus problem is not solvable (without signatures) if a third or more of the processes are faulty~\cite{DBLP:journals/toplas/LamportSP82}. 
\technicalReport{Thus, optimally resilient signature-free Byzantine consensus algorithms, tolerate $t<n/3$ faulty processes.}  
The task is also impossible if a process can impersonate some other process in its communication with the other entities~\cite{DBLP:conf/podc/Ben-Or83}. 
\ems{As in~\cite{DBLP:conf/podc/MostefaouiMR14}, we} assume the absence of spoofing attacks and similar means of impersonation.
}

\Subsubsection{Non-self-stabilizing BFT \ems{consensus}}
\label{sec:nonSSBFTsols}
Rabin~\cite{DBLP:conf/focs/Rabin83} offers a solution to BFT \emph{consensus} (cf. \SectionAbv~\ref{sec:recyclableGMRS} for definitions). 
It assumes the availability of \ems{\emph{random common coins} (RCCs),} allowing for a polynomial number of communication steps and optimal resilience, \ie $t<n/3$, \ems{where $n$ is the number of participating processes.} 
Mostéfaoui, Moumen, and Raynal~\cite{DBLP:conf/podc/MostefaouiMR14}, or MMR in short, is a signature-free BFT binary consensus solution.
MMR is optimal in resilience, uses $O(n^2)$ messages per consensus invocation, and completes within $O(1)$ expected time.
	
\Subsubsection{Non-self-stabilizing synchronous BFT multivalued consensus}
The proposed recycling mechanism uses an SSBFT multivalued consensus, which is based on a non-self-stabilizing BFT multivalued consensus. 
Kowalski and Most{\'{e}}faoui~\cite{DBLP:conf/podc/KowalskiM13} proposed the first multivalued optimal resilience, polynomial communication cost, and optimal $t+1$ rounds, but without early stopping. Abraham and Dolev~\cite{DBLP:conf/stoc/AbrahamD15} advanced the state of the art by offering also optimal early stopping.
Unlike the above BFT\reduce{ multivalued consensus} solutions, our SSBFT multivalued solution \ems{adds} self-stabilization.
	
\technicalReport{
\Subsubsection{SSBFT solutions} 
%
}
In the broader context of SSBFT solutions for message-passing systems, we find topology discovery~\cite{DBLP:conf/netys/DolevLS13}, storage~\cite{DBLP:conf/sss/BonomiPPT18,DBLP:conf/ipps/BonomiPT15,DBLP:conf/podc/BonomiDPR15}, clock synchronization~\cite{DBLP:conf/podc/DolevW95,DBLP:journals/jacm/LenzenR19}, approximate agreement \cite{DBLP:journals/tcs/BonomiPPT19}, asynchronous unison~\cite{DBLP:journals/jpdc/DuboisPNT12}, communication in dynamic networks~\cite{DBLP:conf/opodis/Maurer20}, and SSBFT state-machine replication~\cite{DBLP:conf/sss/BinunCDKLPYY16,DBLP:conf/cscml/DolevGMS18} to name a few.

\Subsubsection{\ems{SSBFT consensus}}
To the best of our knowledge, the only SSBFT RCCs construction is the one by Ben{-}Or, Dolev, and Hoch~\cite{DBLP:conf/podc/Ben-OrDH08}, in short BDH, for synchronous systems with private channels. 
BDH uses its SSBFT RCCs construction as a building block for devising an SSBFT clock synchronization solution.
\reduce{Our work borrows several mechanisms from BDH, such as SSBFT RCCs and SSBFT clock synchronization.} 
Recently, Georgiou, Marcoullis, Raynal, and Schiller~\cite{DBLP:conf/netys/GeorgiouMRS21}, or GMRS in short, presented an SSBFT variation on MMR, which offers a BFT binary consensus\technicalReport{ solution}.
GMRS preserves MMR's optimality\reduce{ properties}, and thus, we\extendedAbstract{ use}\technicalReport{ base our example consensus object on} GMRS \ems{in this work}.

\emsA{GMRS follows the design criteria of loosely self-stabilizing systems~\cite{DBLP:journals/tcs/SudoNYOKM12}, ensuring task completion but with rare safety violation events. In the context of the studied problem, the former guarantee renders the latter one irrelevant.}
We point out that related work to loosely self-stabilizing systems include randomized congestion control~\cite{DBLP:conf/sss/0001GS19} and leader election~\cite{DBLP:journals/ieicet/SudoOKM20,DBLP:journals/tpds/SudoOKMDL19,DBLP:journals/tcs/SudoOKMDL20,DBLP:conf/sirocco/Izumi15}.

\Subsubsection{Self-stabilizing non-Byzantine fault-tolerant solutions} 
Lundstr{\"{o}}m, Raynal, and Schiller~\cite{DBLP:conf/icdcn/LundstromRS21} presented the first self-stabilizing solution for the problem of binary consensus for message-passing systems where nodes may fail by crashing. 
They provided a line of self-stabilizing solutions~\cite{DBLP:journals/corr/abs-2104-03129,DBLP:conf/icdcs/LundstromRS20,DBLP:conf/netys/LundstromRS20,DBLP:conf/netys/GeorgiouLS19,DBLP:conf/netys/GeorgiouGLS19}. 
This line follows the approach proposed by Dolev, Petig, and Schiller~\cite{DBLP:journals/corr/abs-1806-03498,DBLP:conf/podc/DolevPS15} for self-stabilization in the presence of seldom fairness. 
Namely, in the absence of transient-faults, these self-stabilizing solutions are wait-free and no assumptions are made regarding the system's synchrony or fairness of its scheduler. However, the recovery from transient faults does require fair execution, \eg to perform a global reset, see~\cite{DBLP:conf/netys/GeorgiouGLS19}, but only during the recovery period. 
The studied problem is more challenging than the above due to the presence of Byzantine processes and transient faults.   
Thus, we consider synchrony assumptions.

\Subsection{Our contribution} 
We propose an important building block for reliable distributed systems: a \ems{new} SSBFT mechanism  for recycling SSBFT consensus objects.
The proposed mechanism \ems{stabilizes} within expected $\bigO(\kappa)$ synchronous rounds, where $\kappa \in \bigO(t)$ is a predefined constant (that depends on synchrony assumptions) and $t$ is an upper bound on the number of Byzantine processes.
We also present, to the best of our knowledge, the \ems{first} SSBFT synchronous multivalued consensus solution.
The \ems{novel} composition of (i) SSBFT recycling and (ii) SSBFT \ems{recyclable} objects has a long line of applications, such as replication and blockchain. 
\ems{Thus, our transformation advances the state of the art by facilitating} solutions that are more fault-tolerant than the existing implementations, which cannot recover \ems{from} transient faults.

For convenience, a Glossary is provided in Table~\ref{fig:Glossary}.

\begin{table*}[t!]
	\begin{center}
			\begin{tabular}{|l|l|}
				\hline
				\textbf{Notation} & \textbf{Meaning} \\ \hline \hline
				BC & Byzantine-tolerant Consensus\\ \hline
				BDH &     Ben{-}Or, Dolev, and Hoch~\cite{DBLP:conf/podc/Ben-OrDH08}    \\ \hline
				BFT &    non-self-stabilizing Byzantine fault-tolerant solutions     \\ \hline
				COR & Consensus Object Recycling (Definition~\ref{def:probDef})  \\ \hline
				{DPS} & Dolev, Petig, and Schiller~\cite{DBLP:journals/algorithmica/DolevPS23}     \\ \hline 
				MMR & Mostéfaoui, Moumen, and Raynal~\cite{DBLP:conf/podc/MostefaouiMR14}        \\ \hline
				RCCs&  random common coins   \\ \hline
				SSBFT &    self-stabilizing Byzantine fault-tolerant     \\ \hline
				$\kappa\mathit{-}SGC$&    $\kappa$-state global clock     \\ \hline
			\end{tabular}
			\FF
			\caption{\label{fig:Glossary}Glossary}
	\end{center}
\end{table*}

\begin{algorithm*}[t!]
	\begin{\algSize}
		
		\noindent \textbf{local variables:}\technicalReport{\\}
		\tcc{the algorithm's local state is defined here.}
		
		\fbox{$delivered[\sP]:=[\false,\ldots,\false]$ delivery indications; $delivered[i]$ stores the local} 
		
		\fbox{indication and $delivered[j]$ stores the last received indication from $p_j \in \sP$\;} 
		
		\smallskip
		
		\textbf{constants:} \technicalReport{\\}
		$\mathit{initState} := (\bullet,$ \fbox{$[\false,\ldots,\false])$\;}
		
		\smallskip
		
		\textbf{interfaces:}\technicalReport{\\}  	
		\fbox{\label{ln:recycle}$\mathsf{recycle}()$ \textbf{do} (local state, $delivered)$ $ \gets \mathit{initState}$\;} \tcc{also initialize all attached communication channels~\cite[Ch. 3.1]{DBLP:books/mit/Dolev2000}}
		
		\smallskip
		
		\fbox{$\mathsf{wasDelivered}()$ \textbf{do} \{\textbf{if} $\exists S \subseteq\sP: n\mathit{-}t\leq |S|: \forall {p_k \in S} : delivered[k]=\true$ \textbf{then}} \fbox{\textbf{return} $1$ \textbf{else return} $0$;\}\label{ln:wasDelivered}}
		
		\smallskip
		
		\textbf{operations:} $\mathsf{propose}(v)$ \label{ln:RSproposeV}\textbf{do} $\{$implement the algorithm logic$\}$\;
		
		\smallskip
		
		$\mathsf{result}()$ \label{ln:RresultV}\textbf{do} \Begin{
			\lIf{there is a decided value}{\fbox{\{$delivered[i]\gets \true$;} \Return{$v$}\label{ln:nonBot1}\}} 
			\lElseIf{an error occurred}{\fbox{\{$delivered[i]\gets \true$;} \Return{\blitza}\label{ln:nonBot2}\}}
			\textbf{else return} {$\bot$}\label{ln:RresultVBot};
		}

		\smallskip
		
		\textbf{do forever} \Begin{

			\fbox{\lIf{$\mathsf{result}()=\bot$}{$delivered[i]\gets \bot$}\label{ln:ifResultBotDeliveredBot}\tcc{consistency test}}
			
			\tcc{implementation of the algorithm's logic}
			
			\lForEach{$p_j \in \sP$}{\textbf{send} $\mathrm{EST}(\bullet,$\fbox{$delivered[i])$}  \textbf{to} $p_j$\label{ln:RESTrepeatB}}}			
		
		\textbf{upon} $\mathrm{EST}(\bullet,$\fbox{$delivered\mathit{J})$} \label{ln:RuponEst} \textbf{arrival from} $p_j$ \Begin{ 
			
			$delivered[j]\gets delivered\mathit{J}$\label{ln:RRESTrepeatB}\;
			
			\tcc{merge arriving information with the local one}
			
		}
		

		
		\caption{\label{alg:RconsensusB}A recyclable variation on GMRS; \technicalReport{code for node $p_i$}\extendedAbstract{$p_i$'s code}.}
	\end{\algSize}
\end{algorithm*}

\Section{\ems{Basic Result:} Recyclable SSBFT Consensus Objects}
\label{sec:recyclableGMRS}
\ems{In this section, we define a recyclable variation on the consensus problem, which facilitates the use of an unbounded number of consensus instances via the reuse of a constant number of objects (as presented in \SectionAbv~\ref{sec:bck}). 
Then, we sketch \Algorithm~\ref{alg:RconsensusB}, which presents a recyclable variation on an SSBFT asynchronous consensus algorithm (such as GMRS).
Towards the end of this section, Theorem~\ref{thm:COR-proposal-validity} demonstrates that \Algorithm~\ref{alg:RconsensusB} constructs recyclable SSBFT consensus objects.}


\Subsection{\ems{Byzantine-tolerant Consensus (BC)}}
\ems{This problem} requires agreeing on a\technicalReport{ single} value from a given set $V$, \ems{which every \emsA{(non-faulty)} node inputs via $\mathsf{propose}()$.
It} requires 
\emph{BC-validity}, \ie if all non-faulty nodes propose the same value $v\in V,$ only $v$ can be decided, 
\emph{BC-agreement}, \ie no two non-faulty nodes can decide different values, and 
\emph{BC-completion}, \ie all non-faulty nodes decide a value.
When the set, $V$, from which the proposed values are taken is $\{0,1\}$, the problem is called \emph{binary consensus}. Otherwise, it is referred to as \emph{multivalued consensus}.

\Subsection{\ems{Recyclable Consensus Objects}}
\ems{We study systems that implement consensus objects using storage of constant size allocated at program compilation time. 
Since these objects can be instantiated an unbounded number of times, it becomes necessary to reuse the storage once consensus is reached and each non-faulty node has received the object result via $\mathsf{result}()$.}
\ems{To facilitate this, we assume that the object has two meta-statuses: \emph{used} and \emph{unused}. 
The \emph{unused} status represents both objects that were never used and those that are no longer in current use, indicating they are available (for reuse).
Our definition of recyclable objects assumes that the objects implement an interface function called $\mathsf{wasDelivered}()$ that must return $1$ anytime after the result delivery. 
Recycling is triggered by the recycling mechanism (\SectionAbv~\ref{sec:bck}), which invokes $\mathsf{recycle}()$ at each non-faulty node, thereby setting the meta-status of the corresponding consensus object to \emph{unused}.}
\ems{We specify the task of recyclable object construction as one that requires eventual agreement on the value of $\mathsf{wasDelivered}()$. In detail, if a non-faulty node $p_i$ reports delivery (\ie $\mathsf{wasDelivered}_i()=1$), then all non-faulty nodes will eventually report delivery as well.
We clarify that during the recycling process, \ie when at least one non-faulty node invokes $\mathsf{recycle}()$, there is no need to maintain agreement on the values of $\mathsf{wasDelivered}()$.}

\B
\Subsection{\ems{Algorithm} outline}
\label{sec:ourlineAlg}
\Algorithm~\ref{alg:RconsensusB}'s \fbox{boxed} code lines highlight the \ems{code lines relevant to recyclability.}
\emsA{The set of nodes is denoted by $\sP$. We} avoid the restatement of the algorithm, and to focus \ems{on} the parts that matter in this work, the other parts are given in words (cf. \ems{GMRS}~\cite{DBLP:conf/netys/GeorgiouMRS21} for full details). 
The code uses the symbol $\bullet$ to denote any sequence of values.
We assume that the object allows the proposal of $v$ via $\mathsf{propose}(v)$ (line~\ref{ln:RSproposeV}). 
As in $\mathsf{result}()$ (line~\ref{ln:RresultV}), once the consensus algorithm decides, one of the decided value is returned (line~\ref{ln:nonBot1}). 
\ems{Since the algorithm tolerates transient faults,} the object may need to indicate an internal error via the return of the (transient) error symbol, $\blitza$ (line~\ref{ln:nonBot2}). 
In all other cases,\reduce{ \ie as long as no value was decided,} the $\bot$-value is returned (line~\ref{ln:RresultVBot}). 
GMRS uses a do-forever loop that broadcasts the protocol messages (line~\ref{ln:RESTrepeatB}). 
Any node that receives this protocol message, merges the arriving information with the one stored by the local state (line~\ref{ln:RRESTrepeatB}).

\B
\Subsection{Recyclable variation}
\label{sec:recyclableVar}
\Algorithm~\ref{alg:RconsensusB} uses the array $delivered[\sP]$ (initialized to the vector $[\false,\ldots,\false]$) for delivery indications, where $delivered_i[i]:p_i \in \sP$ stores the local indication and $delivered_i[j]:p_i,p_j \in \sP$ stores the indication that was last received from $p_j$. This indication is set to $\true$ whenever $\done()$ returns a non-$\bot$ value (lines~\ref{ln:nonBot1} to~\ref{ln:nonBot2}). \Algorithm~\ref{alg:RconsensusB} updates $delivered[j]$ according to the arriving values from $p_j$ (lines~\ref{ln:RESTrepeatB} and~\ref{ln:RRESTrepeatB}). The interface function $\mathsf{wasDelivered}()$ (line~\ref{ln:wasDelivered}) returns $1$ if at least $n-t$ entries in $delivered[]$ hold $\true$. The interface function $\mathsf{recycle}()$ (line~\ref{ln:recycle}) allows the node to restart its local state w.r.t. \Algorithm~\ref{alg:RconsensusB}.

\technicalReport{Theorem~\ref{thm:COR-proposal-validity} shows that \Algorithm~\ref{alg:RconsensusB} satisfies \ems{the requirements for Recyclable Consensus Objects, which we defined above.} 
Following the definition of the BC problem and GMRS, the theorem assumes that every \emsA{(non-faulty)} node invokes $\mathsf{result}()$ infinitely often.}

\BB\begin{theorem}
	\label{thm:COR-proposal-validity}
	\technicalReport{Suppose that every \emsA{(non-faulty)} node invokes $\mathsf{result}()$ infinitely often.}
	\Algorithm~\ref{alg:RconsensusB} \ems{offers a recyclable asynchronous consensus object.} 
\end{theorem}\B
\renewcommand{\thmcnt}{\ref{thm:COR-proposal-validity}}
\BB

\extendedAbstract{
	\begin{theoremProofSketch}
		If $\exists i \in \Correct:\mathsf{wasDelivered}_i()=1$, then $\mathsf{result}_i()\neq\bot$ (line~\ref{ln:ifResultBotDeliveredBot}). By BC-completion, eventually $\forall j \in \Correct: \mathsf{result}_j()\neq\bot \land \mathsf{wasDelivered}_j()=1$ (lines~\ref{ln:nonBot1} and~\ref{ln:nonBot2}). 
	\end{theoremProofSketch}
} 

\technicalReport{
	\begin{theoremProof}
		Let $R$ be an unbounded execution of \Algorithm~\ref{alg:RconsensusB} in which no \emsA{(non-faulty)} node invokes $\mathsf{recycle}()$.
		Suppose $\exists i\in \Correct:\mathsf{wasDelivered}_i()=1$ in any system state in $R$.
		We show that the system reaches a state $c \in R$ in which $\forall j \in \Correct:\mathsf{wasDelivered}_j()=1$.
		
		By line~\ref{ln:ifResultBotDeliveredBot} and the assumption that $i \in \Correct:\mathsf{wasDelivered}_i()=1$ holds throughout $R$, $\mathsf{result}_i()\neq\bot$ holds in every system state $R$, \ie $p_i$'s state encodes completion.
		By BC-completion and the assumption that \Algorithm~\ref{alg:RconsensusB} is an SSBFT implementation of consensus, $\forall j \in \Correct: \mathsf{result}_j()\neq\bot$ eventually.
		By lines~\ref{ln:nonBot1} to~\ref{ln:nonBot2} and the theorem assumption that every \emsA{(non-faulty)} node invokes $\mathsf{result}()$ infinitely often, $\forall j \in \Correct:\mathsf{wasDelivered}_j()=1$. 
	\end{theoremProof}
} 

\Section{System Settings for the Recycling Mechanism}
\label{sec:sys}
This model considers a synchronous message-passing system. 
The system consists of a set, $\sP$, of $n$ \emph{nodes} (sometimes called \emph{processes} or \emph{processors}) with unique identifiers. 
At most $t < n/3$, out of the $n$ nodes, are faulty. 
Any pair of nodes $p_i,p_j \in \sP$ has access to a bidirectional reliable communication channel, $\mathit{channel}_{j,i}$. 
%
%
	%
	%
	In the \emph{interleaving model}~\cite{DBLP:books/mit/Dolev2000}, the node's program is a sequence of \emph{(atomic) steps}. 
	Each step starts with (i) the communication operation \ems{for receiving messages} that is followed by (ii) an internal computation, and (iii) finishes with a single $send$ operation. 
	The \emph{state}, $s_i$, of node $p_i \in \sP$ includes all of $p_i$'s variables and $\mathit{channel}_{j,i}$. 
	The term \emph{system state}\technicalReport{ (or configuration)} refers to the tuple $c = (s_1, s_2, \cdots,  s_n)$. 
Our model also assumes the availability of a $\kappa$-state global clock, reliable communications, and random common coins (RCCs).

	
	\Subsection{A $\kappa$-state global clock}
	\label{sec:kappaState}
	We assume that the algorithm takes steps according to a common global pulse (beat) that triggers a simultaneous step of every node in the system. 
	Specifically, we denote synchronous executions by $R=c[0],c[1],\ldots$, where $c[x]$ is the system state that immediately precedes the $x$-th global pulse. 
	And, $a_i[x]$ is the step that node $p_i$ takes between $c[x]$ and $c[x+1]$ simultaneously with all other nodes. 
	We also assume that each node has access to a $\kappa$-state global clock via the function $clock(\kappa)$, which returns an integer between $0$ and $\kappa-1$.
	Algorithm 3 of BDH~\cite{DBLP:conf/podc/Ben-OrDH08} offers an SSBFT $\kappa$-state global clock that stabilizes within a constant time.    
	
	\Subsection{Reliable communication}
	\label{sec:relComm}
	Recall that we assume the availability of reliable communication.
	Also, any non-faulty node $p_i \in \sP$ starts any step $a_i[x]$ with receiving all pending messages from all nodes. And, if $p_i$ sends any message during $a_i[x]$, it does so only at the end of $a_i[x]$. We require (i) any message that a non-faulty node $p_i$ sends during step $a_i[x]$ to another non-faulty node $p_j$ is received at $p_j$ at the start of step $a_j[x+1]$, and
	(ii) any message that $p_j$ received during step $a_j[x+1]$, was sent at the end of $a_i[x]$. 
	
	\Subsection{Random common coins (RCCs)}
	\label{sec:rcc}
	As\reduce{ already} mentioned, BDH presented a synchronous SSBFT RCCs solution\reduce{ for message passing systems}. Algorithm $\mathcal{A}$, which has the output of $rand_i \in \{0,1\}$, is said to provide an RCC if $\mathcal{A}$ satisfies the following:
	
	\begin{itemize}[topsep=0.2em, partopsep=0pt, parsep=0pt, itemsep=3pt,leftmargin=3pt, rightmargin=0pt, listparindent=0pt, labelwidth=0.15em, labelsep=0.35em, itemindent=0.75em]
		\item \textbf{RCC-completion:} $\mathcal{A}$ \technicalReport{provides an output}\extendedAbstract{completes} within $\Delta_{\mathcal{A}} \in \mathbb{Z}^+$ synchronous rounds. 
		
		\item \textbf{RCC-unpredictability:} Denote by $E_{x\in \{0,1\}}$ the event that for any non-faulty process, $p_j$, $rand_j=x$ holds with constant probability $p_x > 0$. Suppose either $E_0$ or $E_1$ occurs at the end of round $\Delta_{\mathcal{A}}$. We require that the adversity can predict the output of $\mathcal{A}$ by the end of round $\Delta_{\mathcal{A}}- 1$ with a probability that is not greater than $1 - \min\{p_0, p_1\}$. 
		Following~\cite{DBLP:conf/podc/MostefaouiMR14}, we assume that $p_0=p_1=1/2$.
	\end{itemize}
	
	Our solution depends on the existence of a self-stabilizing RCC service, \eg BDH.
	BDH considers \emph{(progress) enabling} instances of RCCs if there is $x \in \{0,1\}$ such that for any non-faulty process $p_i$, we have $rand_i=x$. BDH correctness proof depends on the consecutive existence of two enabling RCCs instances. 
	
	\Subsection{Legal executions}
	The set of \emph{legal executions} ($LE$) refers to all the executions in which the requirements of task $T$ hold. In this work, $T_{\text{recycl}}$ denotes the task of consensus object recycling \ems{(specified in \SectionAbv~\ref{sec:bck}),} and $LE_{\text{recycl}}$ denotes the set of executions in which the system fulfills $T_{\text{recycl}}$'s requirements. 
	
\Subsection{Arbitrary node failures}
\label{sec:arbitraryNodeFaults}
\ems{As explained in \SectionAbv~\ref{sec:intro},} Byzantine faults model any fault in a node including crashes, arbitrary behavior, and malicious behavior~\cite{DBLP:journals/toplas/LamportSP82}. 
For the sake of solvability~\cite{DBLP:journals/toplas/LamportSP82,DBLP:journals/jacm/PeaseSL80,DBLP:conf/podc/Toueg84}, our fault model limits only the number of nodes that can be captured by the adversary. That is, the number, $t$, of Byzantine failure needs to be less than one-third of the number, $n$, of nodes.
The set of non-faulty nodes is denoted by $\Correct$. 
	
	\Subsection{Arbitrary transient-faults}
	\label{sec:arbitraryTransientFaults}
	We consider any temporary violation of the assumptions according to which the system was designed to operate. We refer to these violations and deviations as \emph{arbitrary transient-faults} and assume that they can corrupt the system state arbitrarily (while keeping the program code intact). 
	Our model assumes that the last\technicalReport{ arbitrary} transient fault occurs before the system execution \ems{starts~\cite{DBLP:series/synthesis/2019Altisen,DBLP:books/mit/Dolev2000}.} Also, it leaves the system to start in an arbitrary state.
	
	\Subsection{Self-stabilization}
	\label{sec:Dijkstra}
	An algorithm is \emph{self-stabilizing} \ems{for} the task of $LE$, when every (unbounded) execution $R$ of the algorithm reaches within a finite period a suffix $R_{legal} \in LE$ that is legal. Namely, Dijkstra~\cite{DBLP:journals/cacm/Dijkstra74} requires $\forall R:\exists R': R=R' \circ R_{legal} \land R_{legal} \in LE \land |R'| \in \mathbb{Z}^+$, where the operator $\circ$ denotes that $R=R' \circ R''$ is the concatenation of $R'$ with $R''$. The part of the proof that shows the existence of $R'$ is called the \emph{convergence} (or recovery) proof, and the part that shows that $R_{legal} \in LE$ is called the \emph{closure} proof. 
	\ems{We clarify that once the execution of a self-stabilizing system becomes legal, it stays legal due to the property of closure.}
	The main complexity measure of a self-stabilizing system is \ems{its stabilization time, which is} the length of the recovery period, $R'$, which is counted by the number of its synchronous rounds.

\Section{SSBFT Recycling Mechanism} 
\label{sec:bck}
\ems{We present an SSBFT recycling mechanism for recyclable objects (\SectionAbv~\ref{sec:recyclableGMRS}). 
The mechanism is a service that recycles consensus objects via the invocation of $\mathsf{recycle}()$ by all (non-faulty) nodes.
The coordinated invocation of $\mathsf{recycle}()$ can occur only after the consensus object \ems{has} terminated and the non-faulty nodes have delivered the result, via $\mathsf{result}()$, as indicated by $\mathsf{wasDelivered}()$.}


\Subsection{\ems{Consensus Object Recycling (COR)}}
\ems{Definition~\ref{def:probDef} specifies the COR problem for a single object.
COR-validity-1 is a safety property requiring that $\mathsf{recycle}()$ is invoked only if there was at least one reported delivery by a non-faulty node. 
COR-validity-2 is a liveness property requiring that eventually $\mathsf{recycle}()$ is invoked. 
COR-agreement is a safety property requiring that all non-faulty nodes simultaneously set the object's status to unused.
This allows any node $p_i$ to reuse the object immediately after the return from $\mathsf{recycle}_i()$. 
}

\begin{figure}[t!]
	\begin{center}
		\includegraphics[scale=0.35, clip]{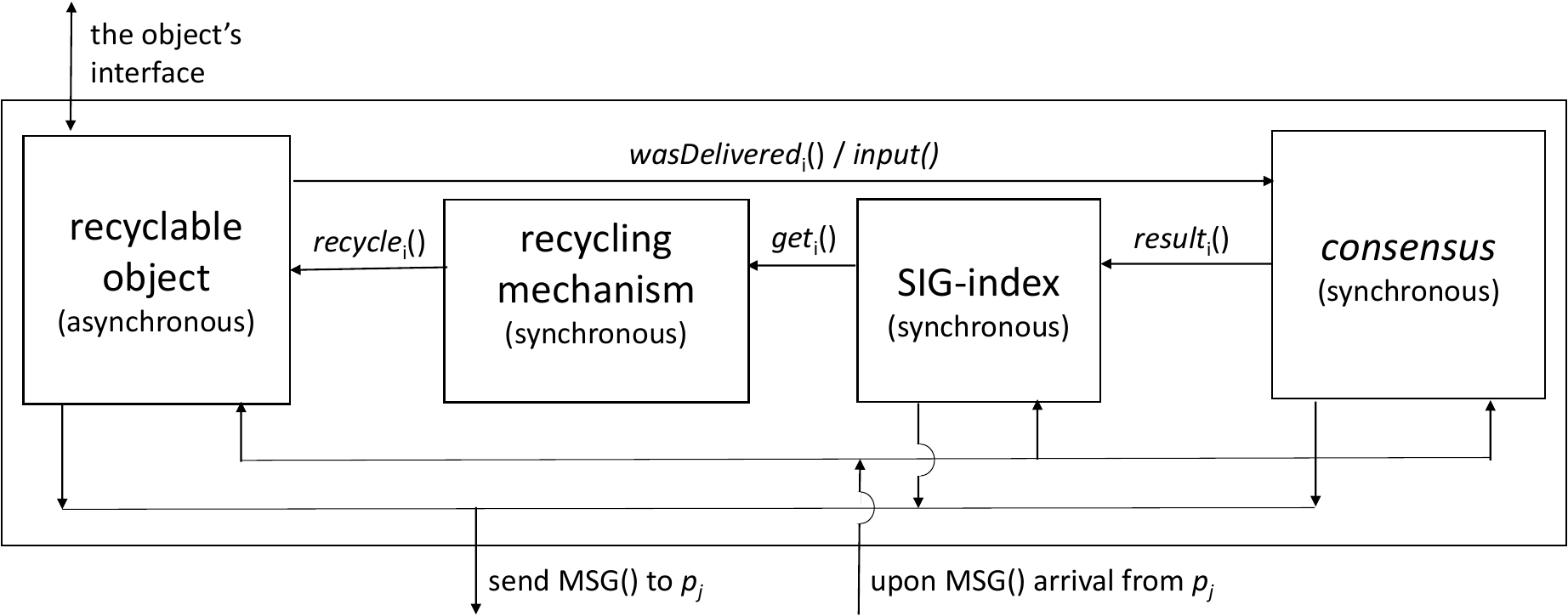}
	\end{center}
		\caption{\label{fig:arc}\small{The solution \ems{uses recyclable objects (\Algorithm~\ref{alg:RconsensusB}), a}  recycling mechanism (\Algorithm~\ref{alg:logOneRecycling}), multivalued consensus \ems{(\Algorithm~\ref{alg:ssbftMVC}), and SIG-index (\Algorithm~\ref{alg:ssbftIndexViaSimRelInc}). \Algorithms~\ref{alg:RconsensusB} and~\ref{alg:ssbftMVC} solve the BC problem (\SectionAbv~\ref{sec:recyclableGMRS}) for asynchronous,  and resp., synchronous settings. \Algorithms~\ref{alg:logOneRecycling} and~\ref{alg:ssbftIndexViaSimRelInc} solve the COR,  resp., SGI-index problems (\SectionAbv~\ref{sec:bck}).}}}
\end{figure}

\B
\begin{definition}[Consensus Object Recycling] 	\technicalReport{The following properties are required:}
	\label{def:probDef}
	\begin{itemize}[topsep=0.2em, partopsep=0pt, parsep=0pt, itemsep=3pt,leftmargin=3pt, rightmargin=0pt, listparindent=0pt, labelwidth=0.15em, labelsep=0.35em, itemindent=0.75em]
		\item \textbf{COR-validity-1:} If a non-faulty node, $p_j$, invokes $\mathsf{recycle}_j()$, then at least one non-faulty node, $p_i$, reported delivery.
		\technicalReport{\Ie no non-faulty node, $p_j$, invokes $\mathsf{recycle}_j()$ if only faulty nodes report delivery.} 
		
		\item \textbf{COR-validity-2:} If all non-faulty nodes report delivery, then at least one non-faulty node, $p_j$, eventually invokes $\mathsf{recycle}_j()$. 
		
		\item \textbf{COR-agreement:} If a non-faulty node invokes $\mathsf{recycle}()$, then all non-faulty nodes, $p_i$, invoke $\mathsf{recycle}_i()$ simultaneously. 
	\end{itemize}
\end{definition}
\BB

\Subsection{\ems{Multiple objects}}
\ems{We also specify that the recycling mechanism makes} sure that, at any time, there are at most a constant number, $logSize$, of active objects, \ie objects that have not completed their tasks. 
Once an object completes its task, the recycling mechanism can allocate a new object by moving to the next array entry as long as some constraints are satisfied. 
Specifically, the proposed solution is based on a synchrony assumption that guarantees that every (correct) node retrieves (at least once) the result of a completed object, $x$, within $logSize$ synchronous rounds since the first time in which at least $t+1$ (correct) nodes have retrieved the result of $x$, and thus, $x$ can be recycled.

\Subsection{Solution overview}
\label{sec:over}
The SSBFT recycling solution is a composition of several algorithms, \ems{see \Figure~\ref{fig:arc}.} Our recycling mechanism is presented in \Algorithm~\ref{alg:logOneRecycling}. It allows every (correct) node to retrieve at least once the result of any object that is stored in a \ems{constant-size} array and yet over time that array can store an unbounded number of object instances. 
\ems{The proposed service mechanism (\Algorithm~\ref{alg:logOneRecycling}) ensures that every instance of the recyclable object, which is implemented by \Algorithm~\ref{alg:RconsensusB}, is guaranteed that every (correct) node calls $\done()$ (line~\ref{ln:wasDelivered}) at least once before all (correct) nodes simultaneously invoke $\mathsf{recycle}()$ (line~\ref{ln:recycle}). This aligns with the solution architecture (\Figure~\ref{fig:arc}).}



We consider the case in which the entity that retrieves the result of object $obj$ might be external (and perhaps, asynchronous) to the proposed solution. The proposed solution does not decide to recycle $obj$ before there is sufficient evidence that, within $logSize$ synchronous cycles, the system is going to reach a state in which $obj$ can be \ems{properly} recycled. Specifically, Assumption~\ref{def:logSize} considers an event that can be locally learned about when $\mathsf{wasDelivered}()$ returns `1' (line~\ref{ln:wasDelivered}).

\begin{algorithm}[t!]
	\begin{\algSize}	
		
		\textbf{constants:} 
		$indexNum$ number of indices of recyclable objects\label{ln:indexNum}\;
		$logSize \in \{0,\ldots,indexNum-2\}$ user-defined bound on the object log size\label{ln:logSize}\;
		
		\smallskip
		
		\textbf{variables:} 
		$obj[indexNum]:$ \extendedAbstract{array of recyclable objects, \eg GMRS}\technicalReport{array of recyclable objects, \eg GMRS. Note that during legal execution only at most $(logSize+1)$ objects are stored at any given point of time}\label{ln:objIndexNum}\; 
		
		\smallskip
		
		$\mathit{ssbftIndex}:$ an SSBFT index of the current object in use (\Algorithm~\ref{alg:ssbftIndexViaSimRelInc})\;
		
		\smallskip	
		
		\textbf{upon pulse} \texttt{/* signal from global pulse system */} \label{ln:vbbBradcastAAA} \Begin{
			\technicalReport{\indent \ForEach{$x \notin \{y ~\bmod$ $ indexNum:y\in \{z-logSize,\ldots,z\}\}$ \textbf{\emph{where}} $z=indexNum+\mathit{ssbftIndex}.getIndex()$}{$obj[x].\mathsf{recycle}()$}}
			\extendedAbstract{\indent \lForEach{$x \notin \{y ~\bmod$ $ indexNum:y\in \{z-logSize,\ldots,z\}\}$ \textbf{\emph{where}} $z=indexNum+\mathit{ssbftIndex}.getIndex()$}{$obj[x].\mathsf{recycle}()$}}
		}
		
		\caption{\label{alg:logOneRecycling}SSBFT synchronous recycling\technicalReport{ with a predefined log size}; \technicalReport{code for node $p_i$}\extendedAbstract{$p_i$'s code}}		
	\end{\algSize}
	
\end{algorithm}

\BB
\begin{assumption}[A bounded time result \emsA{retrieval}]
\label{def:logSize}
Let us consider the system state, $c[r]$, in which the result of object $obj$ was retrieved by at least $t+1$ (correct) nodes. We assume, within $logSize$ synchronous cycles from $c[r]$, the system reaches a state, $c[r+logSize]$, in which all $n-t$ (correct) nodes have retrieved the result of $obj$ at least once.
\end{assumption}
\BB\B

\Algorithm~\ref{alg:logOneRecycling}'s recycling guarantees are facilitated by an SSBFT multivalued consensus object (\Algorithm~\ref{alg:ssbftMVC}).
It helps to decide on a single piece of evidence from all collected ones \ems{(regarding recyclability)} and \Algorithm~\ref{alg:ssbftIndexViaSimRelInc} uses the agreed evidence for updating\reduce{ the value of} the index that points to the current entry in the object array.
We later add details on \Algorithm~\ref{alg:logOneRecycling} before proving its correctness (Theorem~\ref{thm:ssbftBinCon}).

\Subsubsection{Evidence collection using an SSBFT (multivalued) consensus (\Algorithm~\ref{alg:ssbftMVC})}
The SSBFT multivalued consensus protocol returns within $t+1$ synchronous rounds an agreed non-$\bot$ value as long as at least $t+1$ nodes proposed that value, \ie at least one (correct) node proposed that value. 
As mentioned, $\mathsf{wasDelivered}()$ (line~\ref{ln:wasDelivered}) provides the \ems{input} to this consensus protocol. 
Thus, whenever '1' is decided, at least one (correct) node gets an indication from at least $n-t$ nodes that they have retrieved the results of the current object. 
This implies that by at least $t+1$ (correct) nodes have retrieved the results, and, by Assumption~\ref{def:logSize}, all $n-t$ (correct) nodes will retrieve the object result within a known number of synchronous rounds. 
Then, the object could be recycled. 
We later add details on \Algorithm~\ref{alg:ssbftMVC} before proving its correctness\technicalReport{ (Theorem~\ref{thm:algorithmMVC})}.


\Subsubsection{SSBFT simultaneous increment-or-get index (SIG-index)}
\Algorithm~\ref{alg:ssbftIndexViaSimRelInc} allows the proposed solution to keep track of the current object index that is currently used as well as facilitate synchronous increments to the index\reduce{ value}. 
We call this task \emph{simultaneous increment-or-get index} (SIG-index). 
During legal executions of \Algorithm~\ref{alg:ssbftIndexViaSimRelInc}, the (correct) nodes assert their agreement on the index value and update the index according to the result of the agreement on $\mathsf{wasDelivered}()$'s value. 
We later add details on \Algorithm~\ref{alg:ssbftIndexViaSimRelInc} before proving its correctness (Theorem~\ref{thm:isSelf}).


\begin{algorithm}[t!]
	\begin{\algSize}

		\textbf{variables:} $currentResult$ stores the most recent result of $co$\label{ln:varResults}\; 
		
		$co$ a (non-self-stabilizing) BFT (multivalued) consensus object\label{ln:varCo}\;
		
		\smallskip
		
		\textbf{interface required:} 
		
		$input():$ source of (the proposed values) of the given consensus protocol\;
		
		\smallskip
		
		\textbf{interface provided:}

		$\done()$\label{ln:doneSSBFTcon}: \textbf{do} $\Return{(currentResult)}$ \texttt{//} \extendedAbstract{most recent $co$'s decided value}\technicalReport{decided value of the most recent $co$'s invocation}\;
		
		\smallskip
		
		\textbf{message structure:} 
		$\langle appMsg \rangle$, where $appMsg$ is the application message, \ie a message sent by the given consensus protocol\; 
		
		\smallskip
		
		\textbf{upon pulse} \texttt{/* signal from global pulse system */} \label{ln:vbbBradcastAAAB} \Begin{

			\textbf{let} $M$ be \ems{a} message that holds at $M[j]$ the arriving $\langle appMsg_j \rangle$ messages from $p_j$ for the current synchronous round and $M'=[\bot, \ldots,\bot]$\label{ln:messageArrival}\;

			\If{$clock(\kappa)=0$\label{ln:case0co}}{
				
				$currentResult \gets co.\done()$\label{ln:getResults}\;  
				$co.\mathit{restart}()$\label{ln:coRestart}\;				
				$M' \gets co.\mathit{propose}(input())$\label{ln:coPrupuse} \texttt{//} {for recycling $input()\equiv\mathsf{wasDelivered}()$\label{ln:coWasDelivered}}
				
			}
			
			\lElseIf{$clock(cycleSize) \in \{1,\ldots,t\}$\label{ln:processIf}}{$M' \gets co.\mathit{process}(M)$\label{ln:processThen}}
			
			\lForEach{$p_j \in \sP$\label{ln:sendMjToAllIFor}}{\textbf{send} $\langle M'[j] \rangle$ \textbf{to} $p_j$\label{ln:sendMjToAllIDo}}
			
		}


		\caption{\label{alg:ssbftMVC}SSBFT \ems{synchronous} multivalued consensus; \technicalReport{code for node $p_i$}\extendedAbstract{$p_i$'s code}}		
	\end{\algSize}
	
\end{algorithm}


\Subsubsection{Scheduling strategy}
As mentioned, our SSBFT multivalued consensus requires $t+1$ synchronous rounds to complete and provide input to \Algorithm~\ref{alg:ssbftIndexViaSimRelInc} and $\kappa-(t+1)$ synchronous rounds after that, any (correct) node can recycle the current object (according to the multivalued consensus result), where $\kappa=\max\{t+1,logSize\}$. 
Thus, \Algorithm~\ref{alg:ssbftIndexViaSimRelInc} has to defer its index updates until that time. \Figure~\ref{fig:schdule} depicts this scheduling strategy, which considers the schedule cycle of $\kappa$. 
That is, the SIG-index and multivalued consensus starting points are $0$ and $\kappa-4$, respectively. 
Note that \Algorithm~\ref{alg:logOneRecycling} does not require scheduling since it accesses the index only via \Algorithm~\ref{alg:ssbftIndexViaSimRelInc}'s interface of SIG-index, see \Figure~\ref{fig:arc}.  

\begin{algorithm}[t!]
	\begin{\algSize}	
		
		
		\textbf{constants:} $I:$ bound on the number of states an index may have\; 
		
		\smallskip
		
		\textbf{variables:} 
		$\mathit{index} \in \{0,\ldots, I-1\}:$ a local copy of the global\reduce{ logical} object index\;
		
		\smallskip
		
		$\mathit{ssbftCO}:$ an SSBFT multivalued consensus object (\Algorithm~\ref{alg:ssbftMVC})\technicalReport{ that is} used for agreeing on the recycling state, \ie 1 when there is a need to recycle (otherwise 0)\; 
		
		\smallskip
		
		\textbf{interfaces provided:} 
		$getIndex()$ \textbf{do} \Return{$index$}\;

		\smallskip
		
		\textbf{message structure:} 
		$\langle index \rangle$: the logical object index\; 
		
		\smallskip
		
		\textbf{upon pulse} \texttt{/* signal from global pulse system */} \label{ln:vbbBradcastAAAC} \Begin{
			
			\textbf{let} $M$ be the arriving $\langle index_j\rangle$ messages from $p_j$\label{ln:MarrivingIndex}\;
			
			\Switch{$clock(\kappa)$\label{ln:switchA} \emph{\texttt{/* consider $clock()$ at the pulse beginning */}}}{
				
				\lCase{$\kappa-4$\label{ln:case0}}{\textbf{broadcast} $\langle index=getIndex() \rangle$\label{ln:clockAZero}}
				
				\Case{$\kappa-3$}{
					
					\textbf{let} $propose := \bot$\label{ln:proposeGetsBotprp}\;
					\lIf{$\exists v \neq \bot:|\{  \langle v  \rangle \in M\}|\geq \emsA{n-t}$}{$propose \gets v$\label{ln:existsVNeqBotVBull}}
					
					\textbf{broadcast} $\langle propose \rangle$\label{ln:broadcastPropose}\;
				}
				
				\Case{$\kappa-2$\label{ln:case2}}{
					
					\textbf{let} $bit := 0$\label{ln:saveBitGetsBot}; $save \gets \bot$\label{ln:saveBitGetsBotS}\;
					
					\lIf{$\exists s \neq \bot:|\{  \langle s  \rangle \in M\}|> n/2$}{$save\gets s$\label{ln:sNeqBot}}
					
					\lIf{$|\{\langle save \neq \bot\rangle \in M\}|\geq \emsA{n-t}$}{$bit \gets 1$\label{ln:saveNeqBot}}
					
					\lIf{$save = \bot$}{$save \gets 0$\label{ln:saveBot}}
					
					\textbf{broadcast} $\langle bit\rangle$\label{ln:saveBotX}\;	
				}
				
				\Case{$\kappa-1$\label{ln:case3}}{
					\textbf{let} $\mathit{inc} := 0$\label{ln:glbSvV}\;
					
					\lIf{$\mathit{ssbftCO}.\done()$}{$\mathit{inc}\gets 1$\label{ln:glbSvVONE}}
					
					\lIf{$|\{\langle 1\rangle \in M\}|\geq \emsA{n-t}$}{$\mathit{index} \gets (save + \mathit{inc})\bmod I$\label{ln:logicSavePlusSv}}
					\lElseIf{$|\{\langle 0\rangle \in M\}|\geq \emsA{n-t}$}{$\mathit{index} \gets 0$\label{ln:logicZero}}
					\lElse{$\mathit{index} \gets rand(save + \mathit{inc}) \bmod I$\label{ln:logicRandSavePlusSv}}
				}				
			}
		}

		\caption{\label{alg:ssbftIndexViaSimRelInc}SSBFT \ems{synchronous} SIG-index; $p_i$'s code}		
	\end{\algSize}	
\end{algorithm}

\Subsubsection{Communication piggybacking and multiplexing}
We use a piggybacking technique to facilitate the spread of the result (decision) values of the recyclable objects. As \Figure~\ref{fig:arc} illustrates, all communications are piggybacked. Specifically, we consider a meta-message $MSG()$ that has a field for each message sent by all algorithms in \Figure~\ref{fig:arc}. 
That is, when any of these algorithms is active, its respective field in $MSG()$ includes a non-$\bot$ value. 
With respect to GMRS's field, $MSG()$ includes the most recent message that GMRS has sent (or currently wishes to send). 
This piggybacking technique allows the multiplexing of timed and reliable communication (assumed for \ems{the recycling mechanism)} and fair communication (assumed for \ems{the recyclable object).}

\Subsection{SSBFT recycling\remove{ in $\mathsf{BSMP_{n,t}[t < n/3,(t+1)\mathit{-}SGC]}$} (\Algorithm~\ref{alg:logOneRecycling})}
\label{sec:ssbftBinCon}
As mentioned, \Algorithm~\ref{alg:logOneRecycling} \ems{has} an array, $obj[]$ (line~\ref{ln:objIndexNum}), of $indexNum$ recyclable objects (line~\ref{ln:indexNum}). The array size needs to be larger than $logSize$ (line~\ref{ln:logSize} and Assumption~\ref{def:logSize}). 
%
%
\Algorithm~\ref{alg:logOneRecycling}'s variable set also includes $\mathit{ssbftIndex}$, which is an integer that holds the entry number of the latest object in use. \Algorithm~\ref{alg:logOneRecycling} accesses the agreed current index \ems{via} $\mathit{ssbftIndex}.getIndex()$. 
This lets the code \ems{to} nullify any entry in $obj[]$ that is not $\mathit{ssbftIndex}.getIndex()$ or at most $logSize$ older than $\mathit{ssbftIndex}.getIndex()$.
Theorem~\ref{thm:ssbftBinCon} shows \Algorithm~\ref{alg:logOneRecycling}'s correctness.

\technicalReport{\begin{figure*}[t!]}
	\extendedAbstract{\begin{wrapfigure}{r}{0.45\textwidth}}
		\begin{center}
			\extendedAbstract{\BBB\BB}
			\includegraphics[scale=0.5, clip]{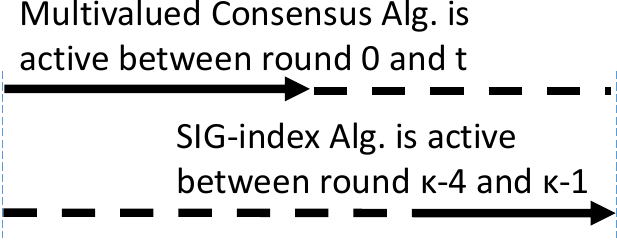}
		\end{center}
		\extendedAbstract{\B}
		\caption{\label{fig:schdule}{The solution schedule uses a cycle of $\kappa=\max\{t+1,logSize\}$ synchronous rounds.}}
		\extendedAbstract{\BBB}
		\extendedAbstract{\end{wrapfigure}}
	\technicalReport{\end{figure*}}


\Subsection{SSBFT \ems{synchronous} multivalued consensus}
\label{sec:ssbftMVC}
\Algorithm~\ref{alg:ssbftMVC} assumes access to a deterministic (non-self-stabilizing) BFT (multivalued) consensus object, $co$, such as the ones proposed by Kowalski and Most{\'{e}}faoui~\cite{DBLP:conf/podc/KowalskiM13} or Abraham and Dolev~\cite{DBLP:conf/stoc/AbrahamD15}, for which completion is guaranteed to occur within $t+1$ synchronous rounds. We list our assumptions regarding the interface to \ems{$co$} in Definition~\ref{sec:assumptionsMVC}.

\Subsubsection{Required \ems{consensus object interface}}
Our solution uses the technique of recomputation of $co$'s floating output~\cite[Ch. 2.8]{DBLP:books/mit/Dolev2000}, where $co$ is specified in Definition~\ref{sec:assumptionsMVC}.		

\BB
\begin{definition}[Synchronous BFT Consensus]
\label{sec:assumptionsMVC}
Let $co$ be a BFT (non-self-stabilizing) synchronous multivalued consensus that implements the following.
\begin{itemize}[topsep=0.2em, partopsep=0pt, parsep=0pt, itemsep=0pt,leftmargin=3pt, rightmargin=0pt, listparindent=0pt, labelwidth=0.15em, labelsep=0.35em, itemindent=0.75em]
\item $\mathit{restart}()$ sets $co$ to its initial state.		

\item $\mathit{propose}(v)$ proposes the value $v$ when invoking (or re-invoking) $co$.\technicalReport{ This operation is effective only after $\mathit{restart}()$ was invoked.} The returned value is a message vector, $M[]$, that includes all the messages, $M[j]$, that $co$ wishes to send to node $p_j$ for the current synchrony round.

\item $\mathit{process}(M)$ runs a single step of $co$.\technicalReport{ The input vector $M$ includes the arriving messages for the current synchronous round, where $M[j]$ is $p_j$'s message.} The returned value is a message vector that includes all the messages that $co$ wishes to send for the current\technicalReport{ synchrony} round.\technicalReport{ This operation is guaranteed to work correctly only after all (correct) nodes have simultaneously taken a consecutive sequence of steps that include invocations of either (i) $\mathit{process}()$, or (ii) $\mathit{restart}()$ immediately before proposing a non-$\bot$ value via the invocation of $\mathit{propose}()$.}

\item $\done()$ returns a non-$\bot$ results after the completion of $co$.\technicalReport{ The returned value is required to satisfy the consensus specifications only if all (correct) nodes have simultaneity taken a sequence of correct $\mathit{process}()$ invocations.}
\end{itemize}
\end{definition}
\BB

\Subsubsection{Detailed description}
\Algorithm~\ref{alg:ssbftMVC}'s set of variables includes $co$ itself (line~\ref{ln:varCo}) and the current version of the result, \ie $currentResult$ (line~\ref{ln:varResults}). This way, the SSBFT version of $co$'s result can be retrieved via a call to $\done()$ (line~\ref{ln:doneSSBFTcon}). \Algorithm~\ref{alg:ssbftMVC} proceeds in synchronous rounds. At the start of any round, node $p_i$ stores all the arriving messages at the message vector $M$ (line~\ref{ln:messageArrival}).

When the clock value is zero (line~\ref{ln:case0co}), it is time to start the re-computation of $co$'s result. Thus, \Algorithm~\ref{alg:ssbftMVC} first stores\reduce{ the current value of} $co$'s result at $currentResult_i$ (line~\ref{ln:getResults}). Then, it restarts $co$'s local state and proposes a new value to $co$ (lines~\ref{ln:coRestart} and~\ref{ln:coPrupuse}). For the recycling solution presented in this paper, the proposed value is retrieved from $\mathsf{wasDelivered}()$ (line~\ref{ln:wasDelivered}).
For the case in which the clock value is not zero (line~\ref{ln:processIf}), \Algorithm~\ref{alg:ssbftMVC} simply lets $co$ process the arriving messages of the current round. 
Both for the case in which the clock value is zero and the case it is not, \Algorithm~\ref{alg:ssbftMVC} broadcasts $co$'s messages for the current round (line~\ref{ln:sendMjToAllIFor}).

\Subsubsection{Correctness proof}
\label{sec:corrAlgorithmMVC}
Theorem~\ref{thm:algorithmMVC} shows that \Algorithm~\ref{alg:ssbftMVC} stabilizes within $2\kappa$\technicalReport{ synchronous} rounds.

\BB\begin{theorem}
\label{thm:algorithmMVC}
\Algorithm~\ref{alg:ssbftMVC} is an SSBFT deterministic (multivalued) consensus solution that \ems{stabilizes} within $2\kappa$ synchronous rounds. 
\end{theorem}\B
\renewcommand{\thmcnt}{\ref{thm:algorithmMVC}}
\BB\begin{theoremProof}
	Let $R$ be an execution of \Algorithm~\ref{alg:ssbftMVC}
	Within $\kappa$ synchronous rounds, the system reaches a state $c \in R$ in which $clock(\kappa)=0$ holds. 
	Immediately after $c$, every (correct) node, $p_i$, simultaneously restarts $co_i$ and proposes the input (lines~\ref{ln:coRestart} and~\ref{ln:coPrupuse}) before sending the needed messages (line~\ref{ln:sendMjToAllIDo}). 
	Then, for the $t<\kappa$ synchronous rounds that follow, all (correct) nodes simultaneously process the arriving messages and send their replies (line~\ref{ln:processIf} and~\ref{ln:sendMjToAllIDo}). 
	Thus, within $2\kappa$ synchronous rounds from $R$'s start, the system reaches a state $c'\in R$ in which $clock(\kappa)=0$ holds. 
	Also, in the following synchronous round, all (correct) nodes store $co$'s results. 
	\ems{These results are}\technicalReport{ guaranteed to be} correct due to Definition~\ref{sec:assumptionsMVC}\technicalReport{'s assumptions}.  	   
\end{theoremProof}\B

\Subsection{SSBFT simultaneous increment-or-get index}
\label{sec:ssbftIndexViaSimRelInc}
The task of \emph{simultaneous increment-or-get index} (SGI-index) requires all (correct) nodes to maintain identical index values that all nodes can independently retrieve via $getIndex()$. 
%
%
The task assumes that all increments are performed according to the result of a consensus object, $\mathit{ssbftCO}$, such as \Algorithm~\ref{alg:ssbftMVC}. \Algorithm~\ref{alg:ssbftIndexViaSimRelInc} presents an SGI-index solution that recovers from disagreement on the index value using RCCs. That is, whenever a (correct) node receives \emsA{$n-t$} reports from other nodes that they have each observed \emsA{$n-t$} identical index values, an agreement on the index value is assumed and the index is incremented according to the most recent result of $\mathit{ssbftCO}$. Otherwise, a randomized strategy is taken for guaranteeing recovery from a disagreement on the index value. Our strategy is inspired by BDH~\cite{DBLP:conf/podc/Ben-OrDH08}\technicalReport{'s SSBFT clock synchronization algorithm}.



\Subsubsection{Detailed description}
\Algorithm~\ref{alg:ssbftIndexViaSimRelInc} is active during four clock phases, \ie $\kappa-4$ to $\kappa-1$. Each phase starts with storing all arriving messages (from the previous round) in the array, $M$ (line~\ref{ln:MarrivingIndex}). The first phase broadcasts the local index value (line~\ref{ln:clockAZero}). The second phase lets each node vote on the majority arriving index value, or $\bot$ in case such value was not received (lines~\ref{ln:proposeGetsBotprp} to~\ref{ln:broadcastPropose}). The third phase resolves the case in which there is an arriving non-$\bot$ value, $save$,  that received sufficient support when voting during phase two (lines~\ref{ln:saveBitGetsBotS} to~\ref{ln:saveBot}). Specifically, if $save\neq \bot$ exists, then $\langle bit=1\rangle$ is broadcast. Otherwise, $\langle bit=0\rangle$ is broadcast. On the fourth phase (lines~\ref{ln:glbSvV} to~\ref{ln:logicRandSavePlusSv}), the (possibly new) index is set either to be the majority-supported index value of phase two plus $inc$ (lines~\ref{ln:glbSvV} to~\ref{ln:logicSavePlusSv}), where $inc$ is the output of $\mathit{ssbftCO}$, or (if there was insufficient support) to a randomly chosen output of the RCC (lines~\ref{ln:logicZero} and~\ref{ln:logicRandSavePlusSv}).

\Subsubsection{Correctness proof}
\label{sec:corrIsSelf}
%
%
%
Theorem~\ref{thm:isSelf} \ems{bounds \Algorithm~\ref{alg:ssbftIndexViaSimRelInc}'s \emsA{stabilization} time.}

\B
\extendedAbstract{
\B\begin{theorem}
	\label{thm:isSelf}
	\Algorithm~\ref{alg:ssbftIndexViaSimRelInc} is an SSBFT SGI-index implementation that stabilizes within expected $\bigO(1)$ synchronous rounds.
\end{theorem}\B
\renewcommand{\thmcnt}{\ref{thm:isSelf}}
\B\begin{theoremProofSketch}
	Lemma~\ref{thm:atLeast} implies that, within $O(1)$ of expected rounds, all (correct) nodes have identical $index$ values. Recall that $c[r] \in R$ is \emph{(progress) enabling} if $\exists x \in \{0,1\}: \forall i \in \Correct: rand_i=x$ holds at $c[r]$ (\SectionAbv~\ref{sec:rcc}).
	Due to the page limit, the Closure proof appears in~\cite{TR}.
	
	\B\B\begin{lemma}[Convergence]
		\label{thm:atLeast}
		Let $r> \kappa$. Suppose $c[r] \in R$ is (progress) enabling system state (\SectionAbv~\ref{sec:rcc}) for which $clock(\kappa) = \kappa - 1$ holds. With probability at least $\min \{p_0, p_1\}$, all (correct) nodes have the same $index$ at $c[r+1]$.
	\end{lemma}\B
	\renewcommand{\lemcnt}{\ref{thm:atLeast}}
	\B\begin{lemmaProofSketch}
		The proof is implied by claims~\ref{thm:atLeastNoVal} to~\ref{thm:atLeastOneVal}.
		
		\B\B\begin{myClaim}
			\label{thm:atLeastNoVal}
			Suppose no (correct) $p_i \in \sP$ receives $\langle x\rangle$ from at least \emsA{$n -t$} different nodes  at $a_i[r]$. 
			With probability $p_0$, any (correct) $p_j $ assigns $0$ to $index_j$ at $a_j[r]$.
		\end{myClaim}\B 
		\renewcommand{\clmcnt}{\ref{thm:atLeastNoVal}}
		\B\begin{claimProof}
			The proof is implied directly from lines~\ref{ln:glbSvVONE} to~\ref{ln:logicRandSavePlusSv}.
		\end{claimProof}\B
		
		\B\begin{myClaim}
			\label{thm:atLeastZeoVal}
			Suppose $p_i : i \in \Correct$ receives $\langle 0\rangle$ from at least \emsA{$n-t$} different nodes at $a_i[r]$. Also, $p_j : j\in \Correct$ receive $\langle x\rangle$ from at least \emsA{$n-t$} different nodes at $a_j[r]$, where $i=j$ may or may not hold. The step $a_j[r]$ assigns $0$ to $index_j$.
			%
		\end{myClaim}\B
		\renewcommand{\clmcnt}{\ref{thm:atLeastZeoVal}}
		\B\begin{claimProofSketch}
			Line~\ref{ln:logicZero} implies the proof since $x=0$.
		\end{claimProofSketch}\B
		
		\B\begin{myClaim}
			\label{thm:atLeastOneValSame}
			Suppose $p_i$ receives $\langle 1\rangle$ from at least \emsA{$n-t$} different nodes at $a_i[r]$. 
			At $c[r]$, $(\mathit{ssbftCO}_i.\done(),save_i)=(\mathit{ssbftCO}_j.\done(),save_j):i,j \in \Correct$.
		\end{myClaim}\B
		\renewcommand{\clmcnt}{\ref{thm:atLeastOneValSame}}
		\B\begin{claimProofSketch}
			At $c[r]$, $\mathit{ssbftCO}_i.\done() = \mathit{ssbftCO}_j.\done()$ holds (BC-agreement).
			There is $p_k :k\in \Correct$ that has sent $\langle 1\rangle$ at $a[r-1]$. By lines~\ref{ln:saveBitGetsBotS} to~\ref{ln:saveNeqBot}, $p_j$ receives at $a_j[r-1]$ the message $\langle x\rangle$ from at least \emsA{$n-t$} different nodes, where $x=save_j \neq \bot$. 
			Any (correct) node broadcasts (line~\ref{ln:broadcastPropose}) either $\bot$ or $x$ at $a[r-2]$. At $a[r-1]$, (correct) nodes receive at most $f <n-2f$ values that are neither $\bot$ nor $x\neq \bot$. Thus, $save_i = save_j$.
		\end{claimProofSketch}\B

		\B\begin{myClaim}
			\label{thm:atLeastOneVal}
			Let $i,j \in \Correct$. Suppose $p_i$ receives $\langle 1\rangle$ from at least \emsA{$n-t$} different nodes at $a_i[r]$ and $p_j$ receives $\langle x\rangle$ from at least \emsA{$n-t$} different nodes at $a_j[r]$. With a probability of at least $\min \{p_0, p_1\}$, $a_i[r]$ and $a_j[r]$ assign the same value to $index_j$, and resp., $index_j$. 
		\end{myClaim}\B
		\renewcommand{\clmcnt}{\ref{thm:atLeastOneVal}}
		\B\begin{claimProofSketch}
			Steps $a[r-1]$ and $a[r]$ independently assign $x$, and resp., $rand$.
			With a probability of at least $\min \{p_0, p_1\}$, all (correct) nodes update $index$ to\reduce{ either} $0$ or $save+inc$ (Claim~\ref{thm:atLeastOneValSame}).
		\end{claimProofSketch}
	\end{lemmaProofSketch}
\end{theoremProofSketch}\B
} 

\technicalReport{
	\B\begin{theorem}
		\label{thm:isSelf}
		Let $R$ be an execution of algorithms~\ref{alg:ssbftMVC} and~\ref{alg:ssbftIndexViaSimRelInc} that is legal w.r.t. \Algorithm~\ref{alg:ssbftMVC} (Theorem~\ref{thm:algorithmMVC}). \Algorithm~\ref{alg:ssbftIndexViaSimRelInc} is an SSBFT SGI-index implementation that stabilizes within expected $\bigO(1)$ synchronous rounds.
	\end{theorem}\B
	\renewcommand{\thmcnt}{\ref{thm:isSelf}}
\B\begin{theoremProof}
Corollaries~\ref{thm:vAvB} and~\ref{thm:atMostOneV} are needed for Lemmas~\ref{thm:atLeast} and~\ref{thm:bmodKappa}. The pigeonhole principle implies Corollary~\ref{thm:vAvB}.


\B\begin{corollary}
	\label{thm:vAvB}
	Let $V_{x \in \{a,b\}}$ be two $n$-length vectors that differ in at most $f<n/3$ entries. For any $x \in \{a,b\}$, suppose $V_x$ contains \emsA{$n-t$} copies of $v_x$. Then $v_a = v_b$.
\end{corollary}\B

Corollary~\ref{thm:atMostOneV} is implied by Corollary~\ref{thm:vAvB}.

\B\begin{corollary}
	\label{thm:atMostOneV}
	Let $c[r] \in R$ be a system state in which $clock(\kappa)=\kappa-3$ and $X=\{x_i:i\in\Correct\}$ be the set of values encoded in the messages $\langle x_i\rangle$ that any (correct) node, $p_i \in \sP$, broadcasts in line~\ref{ln:broadcastPropose} at the end of $a_i[r]$. The set $X$ includes at most one non-$\bot$ value.
\end{corollary}\B

Lemma~\ref{thm:atLeast} implies that, within $O(1)$ of expected rounds, all (correct) nodes have identical $index$ values. Recall that $c[r] \in R$ is \emph{(progress) enabling} if $\exists x \in \{0,1\}: \forall i \in \Correct: rand_i=x$ holds at $c[r]$ (\SectionAbv~\ref{sec:rcc}).

\B\begin{lemma}[Convergence]
	\label{thm:atLeast}
	Let $r> \kappa$. Suppose $c[r] \in R$ is (progress) enabling system state (\SectionAbv~\ref{sec:rcc}) for which $clock(\kappa) = \kappa - 1$ holds. With probability at least $\min \{p_0, p_1\}$, all (correct) nodes have the same $index$ at $c[r+1]$.
\end{lemma}\B
\renewcommand{\lemcnt}{\ref{thm:atLeast}}
\B\begin{lemmaProof}
	The proof is implied by claims~\ref{thm:atLeastNoVal} to~\ref{thm:atLeastOneVal}.
	
	\B\begin{myClaim}
		\label{thm:atLeastNoVal}
		Suppose (i) there is no value $x \in \{0,1\}$ and (ii) there is no (correct) node $p_i \in \sP$ that receives at the start of step $a_i[r]$ the message $\langle x\rangle$ from at least \emsA{$n-t$} different nodes. For any (correct) node, $p_j \in \sP$, it holds that step $a_j[r]$ assigns $0$ to $index_j$ with probability $p_0$.
	\end{myClaim}\B
	\renewcommand{\clmcnt}{\ref{thm:atLeastNoVal}}
	\B\begin{claimProof}
		The proof is implied directly from lines~\ref{ln:glbSvVONE} to~\ref{ln:logicRandSavePlusSv}.
	\end{claimProof}
	
	\B\begin{myClaim}
		\label{thm:atLeastZeoVal}
		Suppose there is a (correct) node $p_i \in \sP$ that receives at the start of step $a_i[r]$ the message $\langle 0\rangle$ from at least \emsA{$n-t$} different nodes. Also, suppose there is $x \in \{0,1\}$ and a (correct) node $p_j \in \sP$ that receives at the start of step $a_j[r]$ the message $\langle x\rangle$ from at least \emsA{$n-t$} different nodes, where $i=j$ may or may not hold. The step $a_j[r]$ assigns $0$ to $index_j$.
		%
	\end{myClaim}\B
	\renewcommand{\clmcnt}{\ref{thm:atLeastZeoVal}}
	\B\begin{claimProof}
		Line~\ref{ln:logicZero} implies the proof since $x=0$ (Corollary~\ref{thm:vAvB}).
	\end{claimProof}
	
	\B\begin{myClaim}
		\label{thm:atLeastOneValSame}
		Suppose there is a (correct) $p_i \in \sP$ that receives at the start of step $a_i[r]$ the message $\langle 1\rangle$ from at least \emsA{$n-t$} different nodes. Let $p_j \in \sP$ be a (correct) node. At $c[r]$, $\mathit{ssbftCO}_i.\done() = \mathit{ssbftCO}_j.\done()$ and $save_i=save_j$ hold.
	\end{myClaim}\B
	\renewcommand{\clmcnt}{\ref{thm:atLeastOneValSame}}
	\B\begin{claimProof}
		At $c[r]$, $\mathit{ssbftCO}_i.\done() = \mathit{ssbftCO}_j.\done()$ holds (\Algorithm~\ref{alg:ssbftMVC}'s agreement property). We show that $save_i=save_j$ holds at $c[r]$.
		Since $p_i$ has received $\langle 1\rangle$ from at least \emsA{$n-t$} different nodes at the start of $a_i[r]$, we know that there is a (correct) node, $p_k \in \sP$, that has sent $\langle 1\rangle$ at the end of $a[r-1]$. By lines~\ref{ln:saveBitGetsBotS} to~\ref{ln:saveNeqBot}, node $p_j$ receives at the start of $a_j[r-1]$ the message $\langle x\rangle$ from at least \emsA{$n-t$} different nodes, where $x=save_j \neq \bot$. 
		By Corollary~\ref{thm:atMostOneV}, any (correct) node broadcasts (line~\ref{ln:broadcastPropose}) either $\bot$ or $x$ at the end of step $a[r-2]$. This means that at the start of $a[r-1]$, (correct) nodes receive at most $f <n-2f$ messages with values that are neither $\bot$ nor $x\neq \bot$. Therefore, $save_i = save_j$ since, at the start of $a_i[r]$ and $a_j[r]$ both $p_i$, and resp., $p_j$ receive from at least \emsA{$n-t$} different nodes the messages  $\langle x_i \rangle$, and resp., $\langle x_j \rangle$, where neither $x_i$ nor $x_j$ is $\bot$.
	\end{claimProof}
	
	\B\begin{myClaim}
	\label{thm:atLeastOneVal}
	Suppose there is a (correct) node $p_i \in \sP$ that receives at the start of step $a_i[r]$ the message $\langle 1\rangle$ from at least \emsA{$n-t$} different nodes. Suppose there is $x \in \{0,1\}$ and a (correct) node $p_j \in \sP$ that receives at the start of step $a_j[r]$ the message $\langle x\rangle$ from at least \emsA{$n-t$} different nodes, where $i=j$ may or may not hold. With a probability of at least $\min \{p_0, p_1\}$, the steps $a_i[r]$ and $a_j[r]$ assign the same value to $index_j$, and resp., $index_j$. 
	%
\end{myClaim}\B
\renewcommand{\clmcnt}{\ref{thm:atLeastOneVal}}
\B\begin{claimProof}
	By Corollary~\ref{thm:vAvB}, $x=1$. The step $a[r-1]$ determines $x$'s value and $rand$ is chosen at the start of step $a[r]$. Due to $rand$'s unpredictability (\SectionAbv~\ref{sec:rcc}), $rand$ and $x$ are two independent values.
	Thus, with a probability of at least $\min \{p_0, p_1\}$, all (correct) nodes update $index$ in the same manner, \ie to either $0$ or $save+inc$ (Claim~\ref{thm:atLeastOneValSame}), where $save$ and $inc$ are values determined by lines~\ref{ln:saveBitGetsBotS} to~\ref{ln:saveBot}, and resp.~\ref{ln:glbSvV} to~\ref{ln:glbSvVONE}.
\end{claimProof}
\end{lemmaProof}

\FF

Lemma~\ref{thm:bmodKappa} shows that all (correct) nodes forever agree on their indexes and simultaneously increment them by one (modulo $I$) only when $clock(\kappa)=\kappa-1$ and $\mathit{ssbftCO}_i.\done()=1$. Lemma~\ref{thm:bmodKappa} uses the following notation. Let $R=c[0],c[1],\ldots,c[r],\ldots$ an unbounded synchronous execution of \Algorithm~\ref{alg:ssbftIndexViaSimRelInc}, where $c[r]$ is the system state that immediately precedes the arrival of the $r$-th common pulse. Denote by $indices^{start}_r$ and $indices^{end}_r$ the sets of all $index_i: i \in \Correct$\remove{ values of (correct) nodes} at $c[r]$, and resp., $c[r+1]$, \ie the beginning, and resp., the end of step $a[r]$. Note that, for all $r$ and $x \in \{start, end\}$, we have $indices^{x}_r \subseteq \{0, 1,\ldots, I -1\}$.

\B\begin{lemma}[Closure] 
	\label{thm:bmodKappa}
	Let $c[r] \in R$, such that $clock(\kappa) = \kappa-1$ at $c[r]$. Suppose $indices^{end}_r = \{v\neq\bot\}$. For every $c[r'] \in R:r'  \in \{r+1,r+\kappa \}$ it holds that $indices^{start}_{r'} = \{ v + x \bmod  I \}$ where $x1$ when $r'=r+\kappa$ and $\mathit{ssbftCO}.\done()=1$. Otherwise, $x=0$, \ie when $r'  \in \{r+1,\ldots ,r+\kappa-1 \}$ or $\mathit{ssbftCO}.\done()\neq 1$.
\end{lemma}\B
\renewcommand{\lemcnt}{\ref{thm:bmodKappa}}
\B\begin{lemmaProof}
	For $r' = r + 1$ the lemma holds since, by definition, $\forall r'': indices^{end}_{r''} = indices^{start}_{r''+1}$. Also, for any system state $c[r']:r' \in \{r+1,\ldots,r +\kappa-1\}$, no (correct) node, $p_i \in \sP$, updates $index_i$ during the step, $a_i[r']$, since $clock(\kappa)\neq\kappa-1$ at $c[r']:r' \in \{r+1,\ldots,r +\kappa-1\}$ and thus lines~\ref{ln:logicSavePlusSv} to~\ref{ln:logicRandSavePlusSv} are not executed, which are the only lines that update $index_i$.
	
	It remains to show that all (correct) nodes, $p_i \in \sP$, update $index_i$ in the same way during the steps $a_i[r']:r'=r+\kappa$ that immediately follow $c[r']$. This is due to the agreement property of \Algorithm~\ref{alg:ssbftMVC}, the arguments above about $c[r']:r' \in \{r+1,r +\kappa-1\}$ as well as Claim~\ref{thm:theSameV}.
	
	\B\begin{myClaim}
		\label{thm:theSameV}
		$indices^{start}_{r+\kappa} = \{v\}:v\neq\bot$.
	\end{myClaim}\B
	\renewcommand{\clmcnt}{\ref{thm:theSameV}}
	\B\begin{claimProof}
		By the schedule (\Figure~\ref{fig:schdule}) and its cycle length, $\kappa$, we know that \Algorithm~\ref{alg:ssbftIndexViaSimRelInc} is not active between $c[r+1]$ and $c[r+\kappa-3]$, but it is active during steps $a[r+\kappa-3]$, $a[r+\kappa-2]$, $a[r+\kappa-1]$, and $a[r+\kappa]$. During steps $a[r+\kappa-3]$, all (correct) nodes broadcast $\langle v \rangle$ (line~\ref{ln:clockAZero}). Thus, at the start of steps $a[r+\kappa-3]$, all (correct) nodes receive $\langle v \rangle$ at least \emsA{$n-t$} times (from different nodes). Thus, during $a[r+\kappa-2]$, all (correct) nodes assign $v$ to their $propose$ variables (line~\ref{ln:existsVNeqBotVBull}) and broadcast  $\langle v \rangle$ (line~\ref{ln:broadcastPropose}). By similar arguments, during $a[r+\kappa-1]$, all (correct) nodes assign $v$ and $1$ to their $save$, and resp., $bit$ variables (lines~\ref{ln:sNeqBot} to~\ref{ln:saveNeqBot}) and broadcast $\langle 1 \rangle$ (line~\ref{ln:saveBotX}). Therefore, all (correct) nodes receive $\langle 1 \rangle$ at least \emsA{$n -t$} times. This implies that during $a[r+\kappa]$, the if-statement condition in line~\ref{ln:logicSavePlusSv} holds and thus $indices^{start}_{r+\kappa} = \{v\neq\bot\}$ holds.
	\end{claimProof}
\end{lemmaProof}
\end{theoremProof}
} 

\B\begin{theorem}
	\label{thm:ssbftBinCon}
	\Algorithm~\ref{alg:logOneRecycling} is an SSBFT recycling mechanism (Definition~\ref{def:probDef})\remove{ for $\mathsf{BSMP_{n,t}[\kappa\mathit{-}SGC,t < n/3,RCCs]}$} that stabilizes within expected $\bigO(\kappa)$ synchronous rounds.
\end{theorem}\B
\renewcommand{\thmcnt}{\ref{thm:ssbftBinCon}}
\B\begin{theoremProof}
%
%
%
\ems{COR-validity-1 and COR-validity-2 are implied by arguments 1 and 2, respectively.}
COR-agreement is implied by Argument 3. 
The stabilization time is due to the underlying algorithms.

\noindent \textbf{Argument 1} \emph{During legal executions, if the value of $index$ is incremented \emph{(}line~\ref{ln:logicSavePlusSv}\emph{)}, $\exists i \in \Correct:\mathsf{wasDelivered}()=1$ holds.~~}
By the assumption that $\mathsf{wasDelivered}()$ provides the proposed values used by the SSBFT multivalued consensus.
\technicalReport{Specifically, \Algorithm~\ref{alg:ssbftMVC} explicitly requires this, see the comment in line~\ref{ln:coPrupuse}.}
The value decided by this SSBFT consensus is used in line~\ref{ln:glbSvVONE} determines whether, during legal executions, the value of $index$ is incremented module $I$ (line~\ref{ln:logicSavePlusSv}), say, from $ind_1$ to $ind_2$.

\noindent \textbf{Argument 2} \emph{During legal executions, if $\forall i \in \Correct:\mathsf{wasDelivered}_i()=1$ holds, $index$ is incremented.~~}
Implied by Argument 1 and BC-validity\technicalReport{ of the consensus protocol}.

\noindent \textbf{Argument 3} \emph{During legal executions, the increment of $index$ is followed by the recycling of a single object, $obj[x]$, the same for all (correct) nodes.~~}
Line~\ref{ln:vbbBradcastAAA} (\Algorithm~\ref{alg:logOneRecycling}) uses the value of $index$ as the returned value from $\mathit{ssbftIndex}.getIndex()$ when calculating the set $S(ind) = \{y ~\bmod indexNum:y\in \{ indexNum + ind-logSize,\ldots, indexNum + ind\}\}$, where $ind \in \{ind_1,ind_2\}$.  
For every $x \notin S(ind)$, $obj[x].\mathsf{recycle}()$ is invoked.
Since $ind_2=ind_1+1 \bmod I$, during legal executions, there is exactly one index, $x$, that is in $S(ind_1)$ but not in $S(ind_2)$. 
\Ie $x=(indexNum + ind_1-logSize) \bmod indexNum$ and only $obj[x]$ is recycled by all (correct) nodes (BC-agreement of the SSBFT consensus).
\end{theoremProof}

\Section{Conclusion}
\label{sec:conclusions}
We have presented an SSBFT algorithm for object recycling.
%
%
Our proposal can support an unbounded sequence of SSBFT object instances.
The\remove{ proposed solution offers optimal fault-tolerance, and the} expected stabilization time is in $\bigO(t)$ synchronous rounds.
We believe that this work is preparing the groundwork needed to construct SSBFT\reduce{ algorithms for} distributed systems, such as Blockchains\reduce{ and the Cloud}.

\emsA{When deploying an asynchronous solution, such as a consensus algorithm, in real-world systems, it is crucial to ensure the solution's correctness remains independent of any timing bounds, which are assumed to be unknown at the time in which the solution is designed and developed. However, given a specific real-world system, which has bounded computation and communication delays, the consensus algorithm can be expected to terminate and deliver results to all non-faulty nodes within a known bounded time, which we refer to as $logSize$ (Assumption~\ref{def:logSize}). This time corresponds to the duration in which it is required to log consensus objects until their results reach all non-faulty nodes. It is important to note that synchrony assumptions are imperative for any deterministic solution to the studied problem since it is equivalent to consensus. This is because the problem entails deciding both the termination of an asynchronous consensus object and whether the agreed-upon value was received by all non-faulty nodes.}    
    
\ems{As a potential avenue for future research, one could explore deterministic recycling mechanisms, say by utilizing the Dolev and Welch approach to SSBFT clock synchronization~\cite{DBLP:journals/jacm/DolevW04}, to design an SSBFT SIG-index. However, their solution has exponential stabilization time, making it unfeasible in practice.}

\subsection*{Acknowledgments}
We express our gratitude to anonymous reviewers for their valuable comments.
The work of E. M. Schiller was partially supported by VINNOVA, the Swedish Governmental Agency
for Innovation Systems through the CyReV project under Grant 2019-03071.

\end{document}